\documentclass{optica-article}

\journal{opticajournal} 

\articletype{Research Article}
\usepackage{here}
\usepackage{lineno}
\usepackage{doi}
\usepackage{siunitx}
\begin{document}

\title{Temperature Dependence of a Depth-Encoded System for Polarization-Sensitive Optical Coherence Tomography using a PM Fiber}
\author{Philipp Tatar-Mathes,\authormark{1,$\dagger$,*}
Rasmus Eilkær Hansen,\authormark{2,$\dagger$}
Samuel Choi \authormark{3},
Manuel J. Marques,\authormark{4}
Niels Møller Israelsen, \authormark{2} 
and Adrian Podoleanu \authormark{4}}
\address{\authormark{$\dagger$} The authors contributed equally to this work.\\
\authormark{1}Optoelectronics Research Center, Physics Unit/ Photonics, Tampere University, Korkeakoulunkatu 3, 33720 Tampere, Finland\\
\authormark{2}DTU Electro, Department of Electrical and Photonics Engineering, Technical University of Denmark, 2800 Kgs.
Lyngby, Denmark \\
\authormark{3}Niigata University, Department of Electrical and Electronics Engineering, 8050 Ikarashi-2, Niigata 950-2181, Japan\\
\authormark{4}Applied Optics Group, School of Physics and Astronomy, Division of Natural Sciences, University of Kent, Canterbury CT2 7NH, United Kingdom}
\email{\authormark{*}philipp.tatar-mathes@tuni.fi} 
\begin{abstract*} 
A polarization-sensitive optical coherence tomography (PS-OCT) system is able to not only show the structure of samples through the analysis of backscattered light, but is also capable of determining their polarimetric properties. This is an extra functionality to OCT which allows the retardance and axis orientation of a bulk sample to be determined. Here, we describe the temperature instabilities of a depth-encoded, multiple input state PS-OCT system, where two waves corresponding to two orthogonal states in the interrogating beam are delayed using a 5-meter long polarization-maintaning (PM) fiber. It is shown that the temperature not only affects the delay between the two relatively delayed waves, but also the amount of mismatched dispersion in the interferometer, which ultimately affects the achievable axial resolution in the system. To this end, the technique of complex master/slave interferometry (CMSI) can be used as an option to mitigate this effect.
\end{abstract*}
\section{Introduction}
Optical coherence tomography (OCT) is a non-destructive imaging technique capable of obtaining depth-resolved images of translucent tissue. Since its first report in 1991 \cite{huang_optical_1991}, OCT has evolved into the \textit{de facto} technology in ophthalmic imaging, and more recently into other medical imaging sub-fields \cite{israelsen_2018}.The first polarization-sensitive OCT (PS-OCT)\cite{hee_polarization-sensitive_1992} system was reported almost immediately after the first OCT system report. This functional extension to standard OCT imaging explores one of the conditions for interference, that of polarization state matching between the waves in the two interferometer arms. By ensuring separate detection of the two orthogonal states, it is possible to ascertain the polarimetric properties of tissues, namely their retardation and optical axis orientation. While the first OCT interferometers were constructed using bulk optical components, nowadays OCT systems are often implemented using fused fiber components due to their ease of alignment and small physical footprint. PS-OCT systems, on the other hand, present some additional challenges which complicate their move into fully fiber-based systems. In particular, the most common implementation of PS-OCT probes the sample with a single, circularly-polarized state \cite{hee_polarization-sensitive_1992};  this is often achieved using bulk polarization optics and free space propagation, as it is necessary to ensure a stable polarization state over time, which is not possible when using single-mode fibers. \cite{marques_polarization-sensitive_2015}. To preserve the orientation of polarization, polarization-maintaining fibers (PM-fiber) 
 can be used, but they present their own set of challenges, one of them being polarization-mode dispersion which can lead to multiple overlapping OCT images, causing ghosting.  If PS-OCT systems are to be used in real practical cases (eg, in the medical practice, especially in endoscopy), then a different approach is needed. A multiple input state configuration such as the one in Wang et al.\cite{wang_depth-encoded_2014} represents a compact solution with a small footprint while being robust to slight misalignments and easy to maintain.
 Using an all-fiber based system is therefore advantageous for clinical adaptations. Here, the visualization of birefringence and collagen alignment \cite{mclean_2019a, li_2019b} can help assess additional tissue contrast when compared to standard OCT. For instance, in the eye, the sclera, retinal nerve fiber layer, and corneal stroma display birefringent properties in the eye. 
But also apart from the eye, PS-OCT has been applied in many biomedical fields including skin imaging \cite{strasswimmer_2004, pircher_2004a, pierce_2004a, yasuno_2002a}, dental imaging \cite{wang_1999a, baumgartner_2000a}, anterior and posterior eye imaging \cite{pircher_2011a, hitzenberger_2001, pircher_transversal_2004, cucu_polarization-sensitive_2004, lim_birefringence_2011, cense_vivo_2002, pircher_imaging_2004, podoleanu_vivo_2004, pircher_human_2006} such as the characterization of atherosclerotic plaque inside blood vessels \cite{nadkarni_measurement_2007, kuo_polarization-sensitive_2007, giattina_assessment_2006}. Recent applications also include the investigation of bronchial airways \cite{james_airway_2009, james_clinical_2007, adams_birefringence_2016, castro_effectiveness_2010, feroldi_vivo_2019, vaselli_polarization_2021, villiger_deep_2016, willemse_vivo_2020}, such as pre-clinical studies for retinal disorders. 
For biologists studying the reproductive capabilities of cells, this information can offer insights into the dynamics by tracking the changes in birefringence over time \cite{zheng_noninvasive_2013, zheng_understanding_2012, zheng_label-free_2012, wang_limited_2001, liu_increased_2000}. 
In order to separate the two orthogonal interrogating polarization states, these are multiplexed in two depth-resolved channels in the OCT signal. While this approach does not require any additional components to separate the channels apart from a polarization-maintaining (PM) fiber, it assumes that the environmental parameters (temperature, mechanical stress etc) are stable enough to ensure consistent channel separation. In this article, we present a study on the temperature dependence of the performance of the PM fiber employed to separate the channels, and its influence on the performance of the PS-OCT system. In particular, we analyse its impact on the axial resolution and the separation between the channels, as well as the dispersion behaviour of the setup for varying temperatures. 
\subsection{Experimental depth-encoded setup}
\begin{figure}[H]
    \centering
    \includegraphics[width=0.9\textwidth]{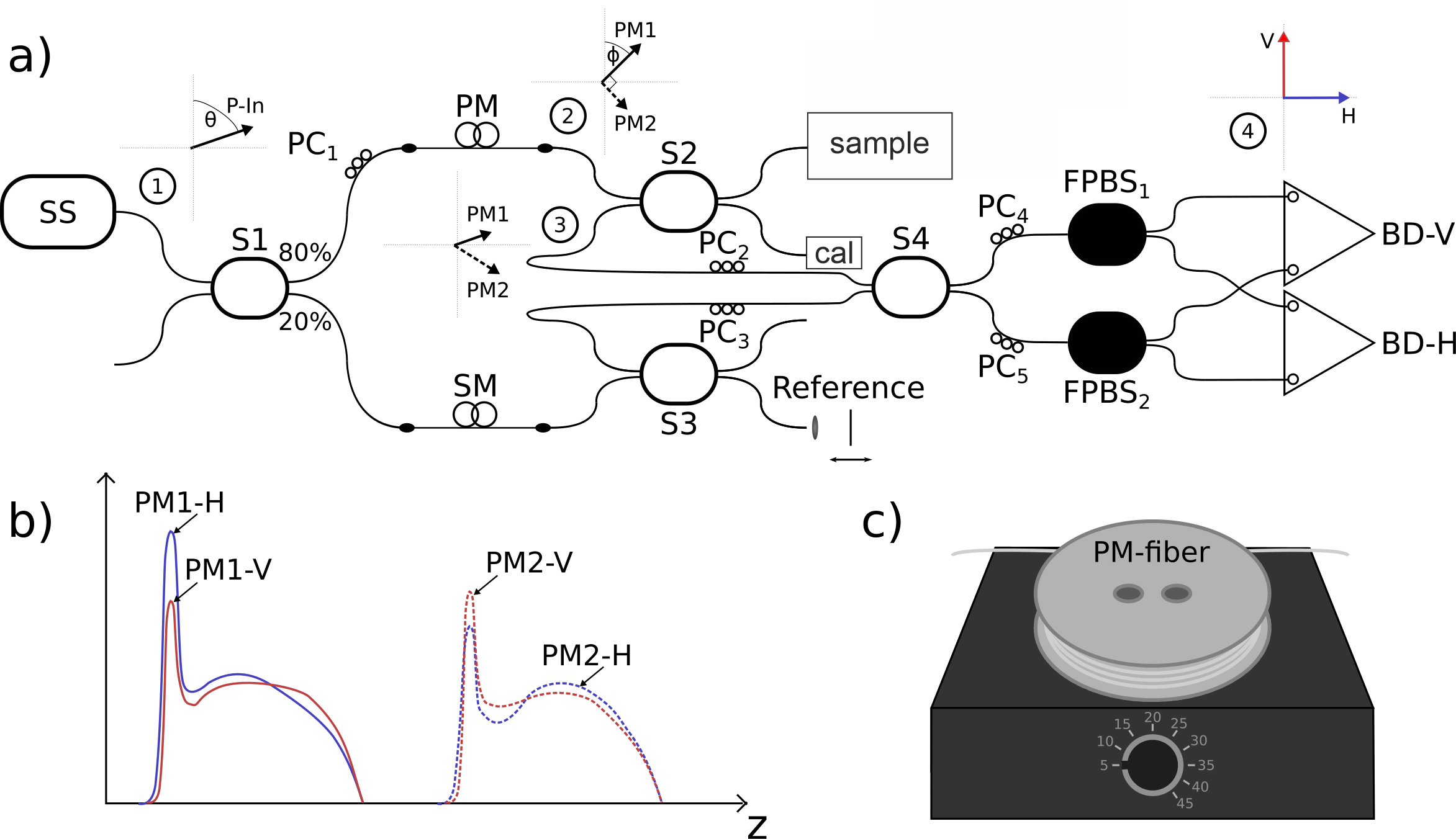}
    \caption{\textbf{a)} A sketch of the PS-OCT setup. Along with the setup, the polarization state of the light is shown in four different positions, numbered accordingly. \textbf{SS} swept source, \textbf{S1-4} splitter 1-4, \textbf{PC1-5} polarization controller 1-5, \textbf{PM} polarization maintaining fiber, \textbf{SM} single mode fiber, \textbf{cal} calibration port, \textbf{FPBS} fiber based polarization beam-splitter, \textbf{BD-V,H} Balanced detection in vertical and horizontal channels, \textbf{P-In} input polarization state, \textbf{PM1,2} polarization states corresponding to fast and slow axes in PM fiber. \textbf{b)} An example of a single A-scan obtained with the setup. The vertical and horizontal channels (blue and red, accordingly) in both channels the two depth resolved signals are measured, here showed in fully drawn and dashed lines according to \textbf{a)}.  \textbf{c)} As the PM fiber is temperature sensitive, it has been attached to a spool which is put in thermal contact to a stabilized heating plate (Thorlabs PTC1), such that the temperature of the PM fiber can be kept constant irrespective of the ambient temperature.}
    \label{fig:setup}
\end{figure}

The experimental setup is a fiber-based PS-OCT setup inspired by Wang et al. \cite{wang_depth-encoded_2014} and is schematically illustrated in Fig. \ref{fig:setup}. 
Light from a swept source, in our case from Axsun, with 100 kHz repetition rate, emitting at \SI{1.3}{\micro \meter} central wavelength with over 100 nm tuning range, is split up into a sample and a reference arm. In the sample arm, a 5 m long PM Panda fiber delays two polarization modes of the interrogating beam that are orthogonal to one another, thus enabling  polarization multiplexing. The distance of the delayed polarization channel corresponds to $( n_2 - n_1) \cdot L $, with $n_1$ and $n_2$ being the corresponding effective refractive indices of the PM fiber for the different polarization modes, and $L$ being the length of the PM fiber. This enables the detection of two polarization modes that are multiplexed into different optical path differences. Both the sample and reference arm have the same length, but the reference arm was implemented using a single-mode (SM) fiber. The interference signal is detected using a custom-made polarization diversity balanced detector unit. It consists of two in-fiber polarization controllers which can be used, in conjunction with the fiber-based polarization beam splitters, to ensure the selection of two orthogonal polarization states from the sample arm. This is performed using the calibration method elaborated in \cite{wang_depth-encoded_2014}.
As a result, four total channels will be obtained that are detected by two balanced photodetector modules (PDB435C-AD (800-1700 nm), Thorlabs) and the signal is digitized using an Alazar ATS9360 12 bit 1.8 GS/s digitizer.
With knowledge of g and h, the matrix T can be computed. With the knowledge of T, the fast Fourier transform of any signal can be calculated without any further data processing being required other than multiplying two matrices. This formalism allows for significant reduction in computation time if information from a few depths only is needed, such as in the case of generating a few \textit{en-face} OCT images only \cite{bradu_2015, chin_masterslave_2016, bradu_masterslave_2016, bradu_recovering_2018}. Before the measurements can be taken, an initial calibration must be carried out. Signal processing for the complex master-slave interferometry (CMSI) was carried out as described in \cite{rivet_complex_2016}. Summarized, the goal of the processing is to gather the previously mentioned functions "$g(\lambda)$" and "$h(\lambda)$". The processing is carried out for both channels by calling out the relevant masks in the corresponding depth ranges. Within our experiments, the same set of experimental masks could be used to generate the complex masks for both channels. The functions are processed for different temperature settings of the PM-fiber. The polarization-sensitive detection is achieved using two fiber-based polarization beam splitters. With the help of matrix calculations \cite{wang_depth-encoded_2014}, it is possible to retrieve the Jones matrix of a sample for every single encodable depth, which ultimately enables the calculation of phase retardance and diattenuation.
The measurement of the polarization state does not directly provide the birefringence and rotation of the sample, but rather of the entire setup. It is demonstrated in \cite{wang_depth-encoded_2014} how the birefringence can still be recovered by referencing the measurement inside the sample to the sample surface. The reflected signal is retrieved using balanced detection. The detector units are calibrated using the same swept source laser as used for the PS-OCT system, which is mostly linearly polarized. We ensured that the resulting output into the calibration port was linearly polarized by inserting a linear polarizer directly after the source. Using this configuration, balanced detection calibration has been performed on both modes with and without a polarizer between the laser and fiber coupling. The intensity is equalized among the two axially separated channels by fine adjustment of PC1, following the detector calibration procedure described in \cite{wang_depth-encoded_2014}. This is done by injecting linearly polarized light into the calibration port of Fig. \ref{fig:setup} and adjusting PC4 and PC5 until only one of the polarization states is detected in the power monitor ports. The best contrast is achieved when there is the same signal in both channels, which corresponds to coupling at 45 degrees with respect to the primary axes of the PM fiber. Any mechanical movement, and changes in the ambient conditions will influence this coupling, therefore it is worth optimising this condition at every start up. We retrieved the signal from the sample surface in two different ways, either by finding the surface of the sample using simple thresholding, or by using the signal of a mirror that is placed prior to introducing the sample under study, and moved along the optical axis to find the corresponding position. Under the assumption that the surface reflection yields a higher signal than the sample at different axial positions, thresholding can be applied to acquire the surface. In a sample with many interfaces this is not generally the case, therefore using a mirror as a sample was found to be the most optimal way of acquiring the surface. When performing depth-resolved measurements, this will generate an extensive amount of data to be calculated. Using CMS, however, vast amounts of computation time can be reduced when taking en-face images, which enables imaging with little latency. Using a mirror as a sample we noticed that the A-scan peak obtained using the complex master/slave formalism fluctuates from day to day along the OPD axis. To calculate the birefringent properties of a sample, a pixel-by-pixel comparison is needed across the 4 channels, which is not possible when the signal is subject to fluctuations. We have identified that the major fluctuations are due to the temperature dependence of the PM fiber parameters. The PM fiber is of the Panda type, which means the birefringence is stress induced. Along the direction of the fiber there are two rods embedded in the cladding with a higher thermal expansion coefficient than the surrounding glass. When the ambient temperature increases the induced tension reduces, thus making the PM fiber less birefringent while increasing its effective length due to heat expansion. The effect of ambient temperature changes in the PM-fiber to the PS-OCT setup is presented in the following section.
 \section{Results}
\begin{figure}[H]
    \centering
    \includegraphics[width=0.8\textwidth]{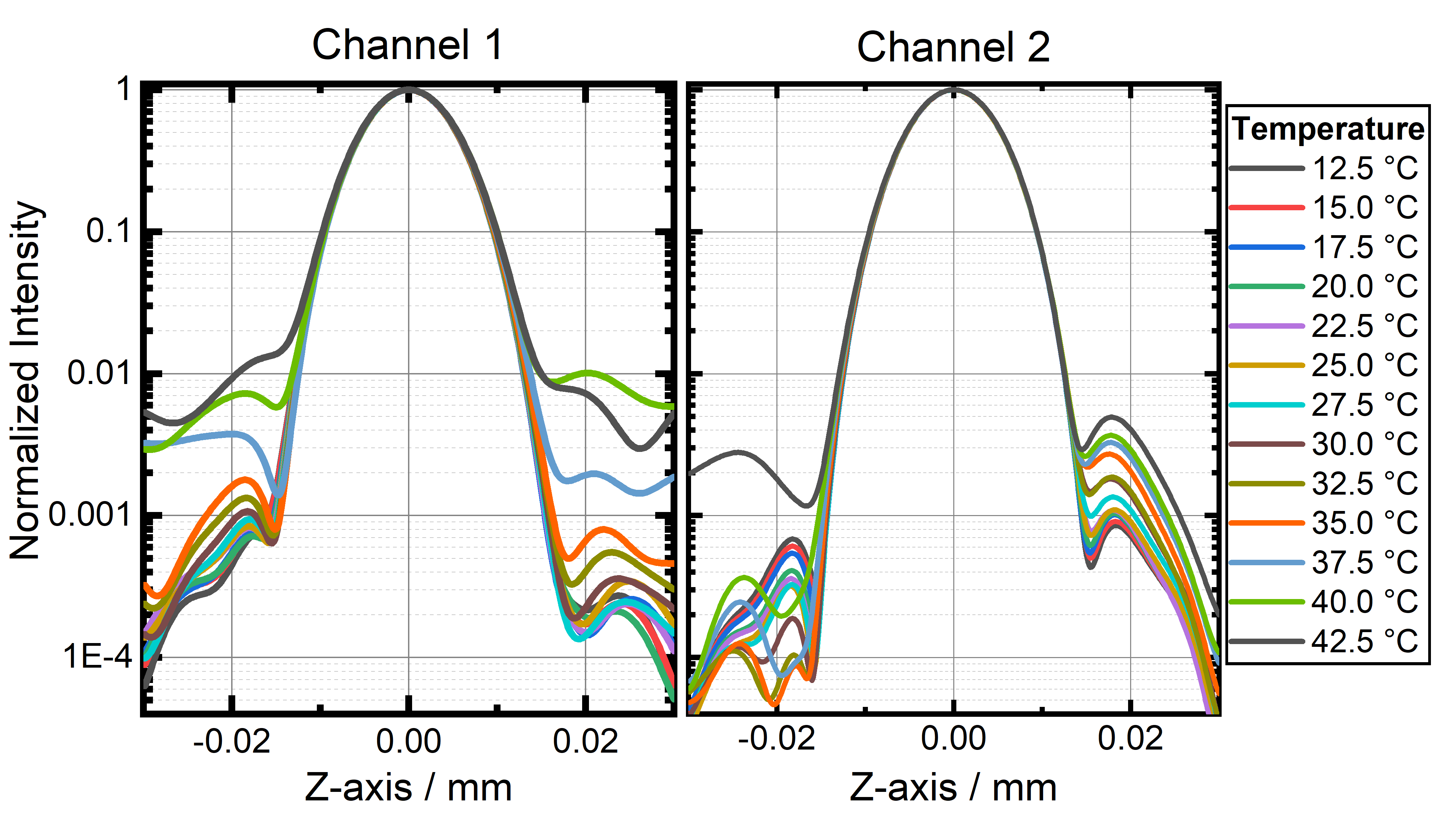}
    \caption{Logarithmic representation of the axial resolution in both channels to illustrate the variation of axial resolution with ambient temperatures of solely the PM-fiber.}
    \label{fig:log1}
\end{figure}

\begin{figure}[H]
    \centering
    \includegraphics[width=0.6\textwidth]{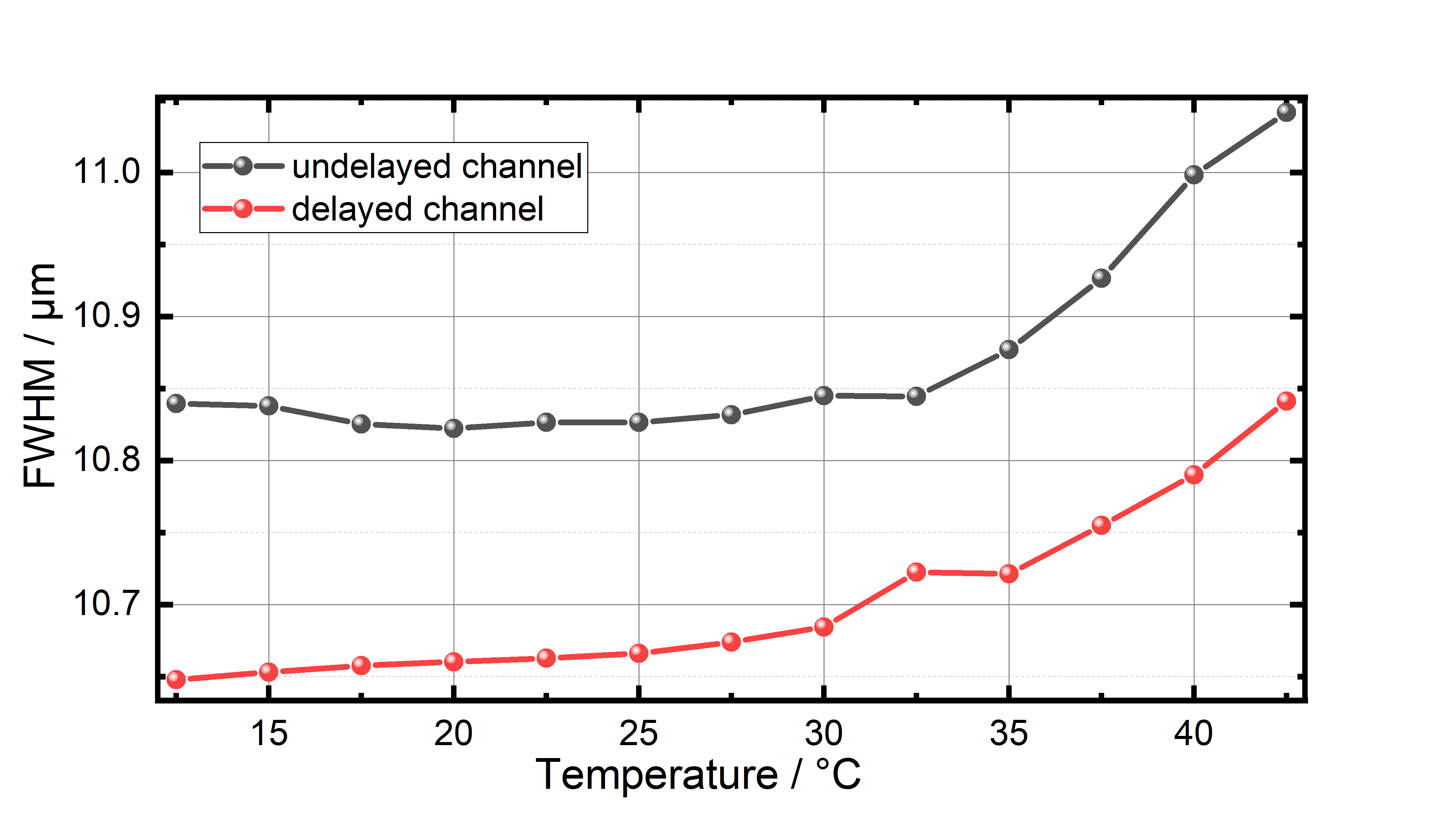}
    \caption{Evaluated FWHM as a function of temperature for the delayed and undelayed channel.}
    \label{fig:fwhm}
\end{figure}

\begin{figure}[H]
    \centering
    \includegraphics[width=0.6\textwidth]{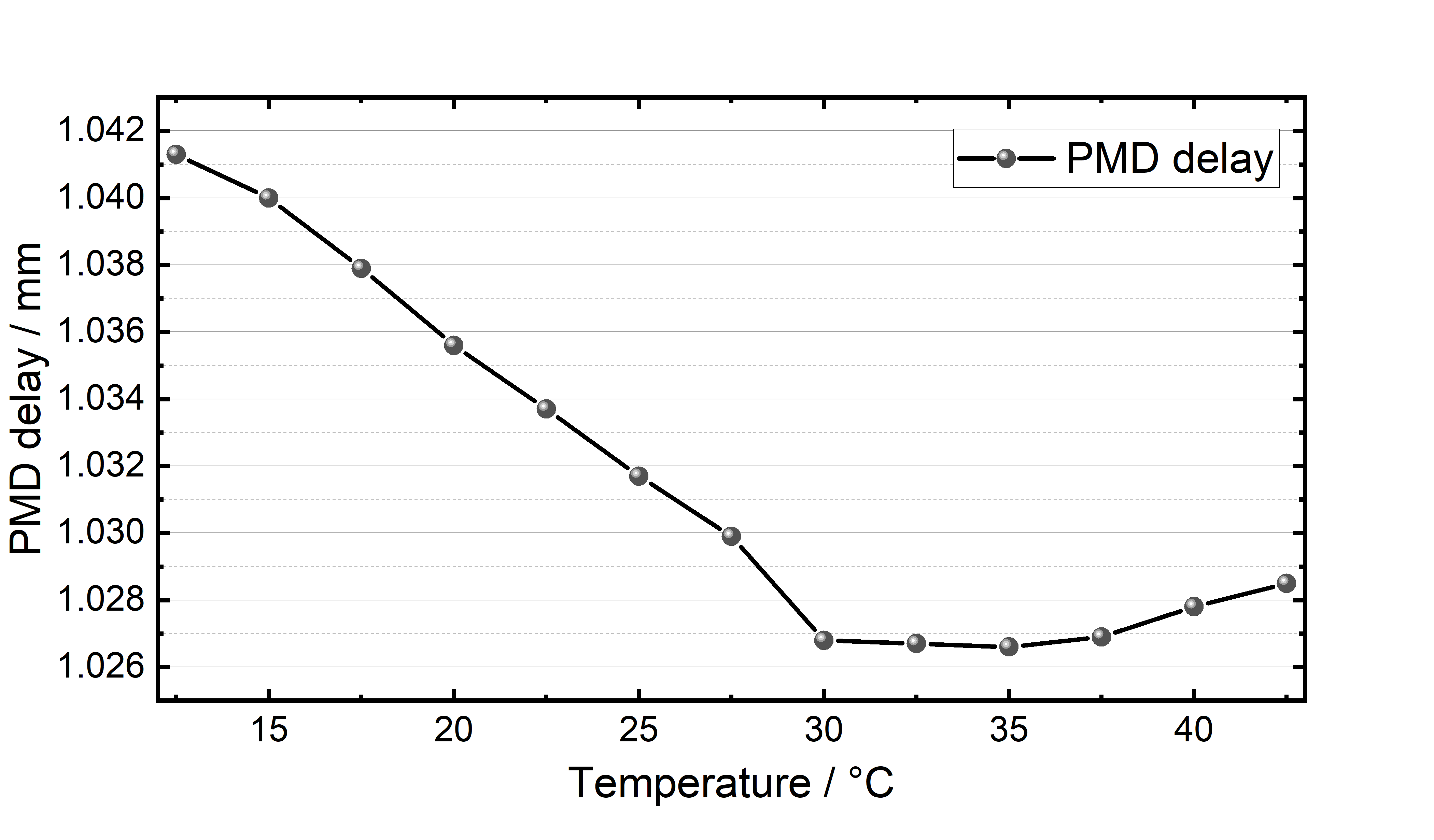}
    \caption{Resulting PMD delay as a function of temperature}
    \label{fig:PMD}
\end{figure}

\begin{figure}[H]
    \centering
    \includegraphics[width=0.6\textwidth]{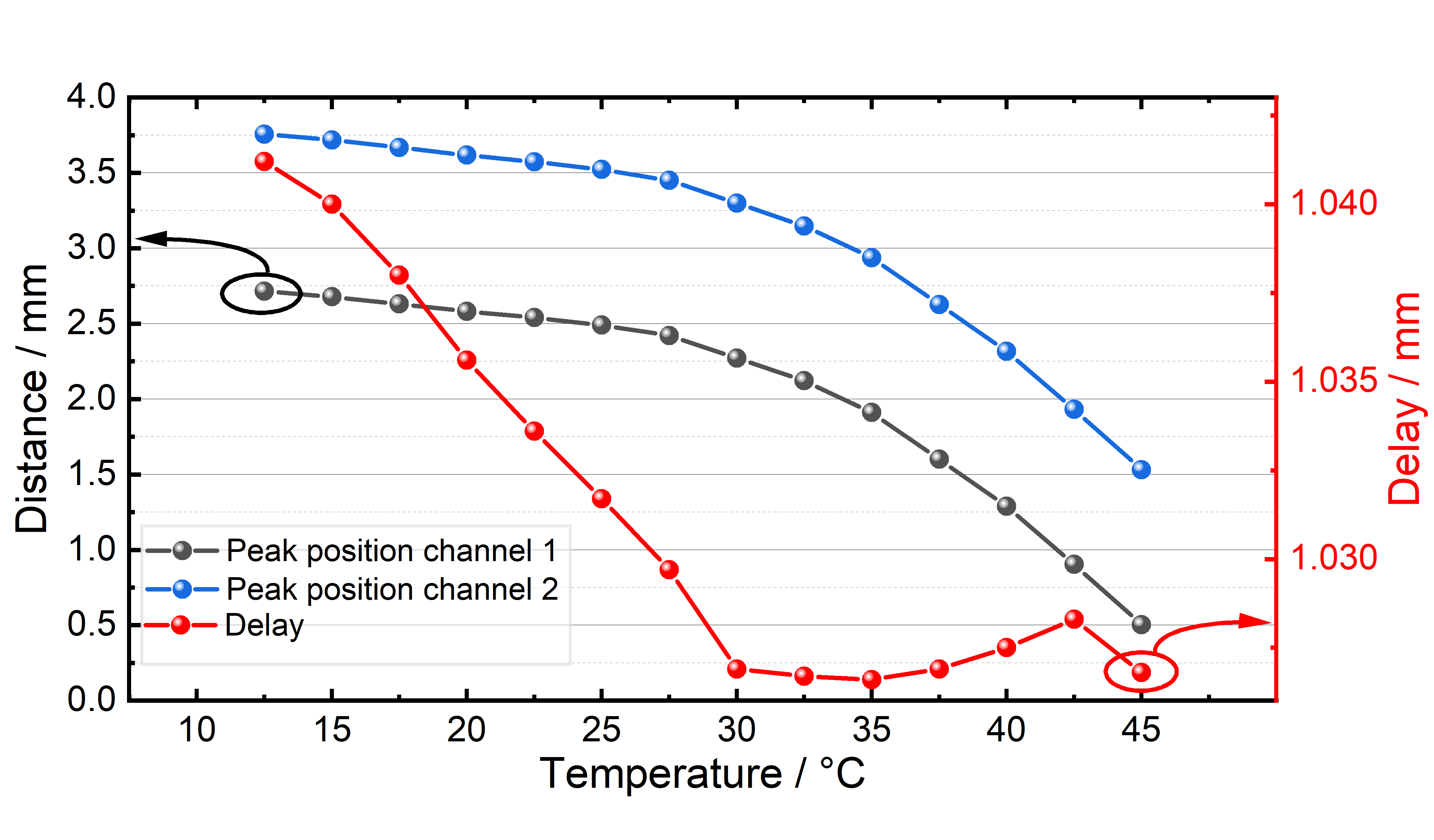}
    \caption{Axial drift with respect to temperature. The delay between both channels is plotted on the right y-axis.}
    \label{fig:drift}
\end{figure}

In our experiments, we ensured that both channels lay within the range allowed by the k-clock of the light source. This way, we could ensure that the variations are solely due to changes of the PM-fiber temperature. We found the setup to be invariant to changes with respect to time. 
In a next step, the axial resolution of both channels were measured - the undelayed (channel 1) and delayed (channel 2) channel - at temperatures ranging from 12.5$~^{\circ}$C to 42.5$~^{\circ}$C. The normalized intensity was plotted logarithmically for both cases. The results can be seen in Fig. \ref{fig:log1}. 
From these graphs, it becomes evident that the pedestals of the peaks are rising with increasing temperature. Furthermore, the effect varies from channel 1 to channel 2. To further analyze this behavior, the FWHM of each peak was evaluated and plotted as a function of temperature. The result can be seen in Fig. \ref{fig:fwhm}. 
It can be concluded that the width of the peaks increases for both channels with rising temperature, leading to a slightly reduced resolution. The change in birefringence of the PM-fiber leads to a broadening of about \SI{0.4}{\micro \meter}  in the FWHM of both channels when varying the temperature by 30$~^{\circ}$C.
Using CMSI, the dispersion of the system was measured for varying temperatures. The result can be seen in Fig.\ref{fig:PMD}.
An increase in temperature leads to an increased amount of unbalanced dispersion to the setup. Simultaneously, the PMD is decreasing at higher temperatures, illustrated by the blue curve in Fig.\ref{fig:PMD}. While varying the temperature, we encountered an additional implication. When the PM-fiber is heated up, its effective length increases. This dependence changes the g and h functions of the system such that the peaks, to which the formalism initially was applied to, will not only deviate from their initial shape, but also from their axial position. Consequently, not only changes in the group velocity dispersion of the fiber occur, but also drifts in the OPD as well as the separation between the 2 channels need to be considered with temperature variation. The resulting polarization mode dispersion (PMD) of the PM-fiber with respect to temperature change is plotted in Fig \ref{fig:PMD}. 
In the resulting  A-scan image, this will appear as a drift of the entire image. The magnitude of the drift and the corresponding peak position of both channels were investigated and are plotted with respect to temperature change in Fig. \ref{fig:drift}.
It can be seen that the variation in the peak position affects both channels equally and leads to a drift of more than 2$~$mm. This drift is so large that it can prevent the algorithm from finding the peak of each corresponding channel and address them accurately, leading to a highly distorted image, or no image at all. In addition to this, the delay between the channels is altered with temperature.In practice, precise knowledge of the fiber behavior and accurate temperature control need to be carefully considered when aiming for stable performance during measurement series using an all-fiber based depth-encoded PS-OCT system over a long period of time. 
\section{Conclusion}
We report on the first complete characterization of dispersion-attributed effects that lead to a degrading imaging performance in all-fiber based PS-OCT setups. Variations in the temperature lead to a change in effective fiber length of the PM-fiber which causes a drift in the system and changes the OPD. In addition to that, it was demonstrated that the polarization mode dispersion decreases with increasing temperature, which leads to a broadening in the detection of both delayed and undelayed channel accuracy. Using a temperature controller, as well as the k-clock provided by the light source, we successfully stabilised variations arising from changes in temperature and characterized them. We have also found that the reported behavior seems to be stable with regards to changes over time. We hope that these findings can be highly beneficial for future depth-encoded, fiber-based PS-OCT setups as this approach represents a simpler, more compact and robust solution thank bulk components, making it feasible for potential clinical imaging applications.
\section{Backmatter}
\begin{backmatter}
\bmsection{Funding}
The work was supported by the European Union's Horizon 2020 research and innovation programme under the Marie Skłodowska-Curie grant agreement No 860807, as well as from the NIHR202879 grant. The work of Adrian Podoleanu was supperted by the UCL institute of Ophthalmology - Moorfields Eye Hospital under grant No BRC003.
\bmsection{Acknowledgments}
\end{backmatter}
\bibliography{manuscript}

\end{document}